\documentclass[11pt]{article}
\usepackage{graphicx}
\usepackage{amssymb,amsmath,amsfonts}

\def\Re{{\cal R \mskip-4mu \lower.1ex \hbox{\it e}\,}}
\def\Im{{\cal I \mskip-5mu \lower.1ex \hbox{\it m}\,}}
\def\ie{{\it i.e.}}
\def\eg{{\it e.g.}}

\def\sub#1{_{\lower.25ex\hbox{$\scriptstyle#1$}}}
\def\tev{\,{\ifmmode\mathrm {TeV}\else TeV\fi}}
\def\gev{\,{\ifmmode\mathrm {GeV}\else GeV\fi}}
\def\mev{\,{\ifmmode\mathrm {MeV}\else MeV\fi}}
\def\mpl{\ifmmode M_{pl}\else $M_{pl}$\fi}
\def\mpl{\ifmmode \overline M_{Pl}\else $\bar M_{Pl}$\fi}
\def\to{\rightarrow}

\def\subw{_{\rm w}}
\def\mh{\ifmmode m\sbl H \else $m\sbl H$\fi}
\def\mch{\ifmmode m_{H^\pm} \else $m_{H^\pm}$\fi}
\def\mt{\ifmmode m_t\else $m_t$\fi}
\def\mc{\ifmmode m_c\else $m_c$\fi}
\def\mz{\ifmmode M_Z\else $M_Z$\fi}
\def\mw{\ifmmode M_W\else $M_W$\fi}
\def\mws{\ifmmode M_W^2 \else $M_W^2$\fi}
\def\mhs{\ifmmode m_H^2 \else $m_H^2$\fi}   
\def\mzs{\ifmmode M_Z^2 \else $M_Z^2$\fi}
\def\mts{\ifmmode m_t^2 \else $m_t^2$\fi}
\def\mcs{\ifmmode m_c^2 \else $m_c^2$\fi}
\def\mchs{\ifmmode m_{H^\pm}^2 \else $m_{H^\pm}^2$\fi}
\def\ztwo{\ifmmode Z_2\else $Z_2$\fi}
\def\zone{\ifmmode Z_1\else $Z_1$\fi}
\def\mtwo{\ifmmode M_2\else $M_2$\fi}
\def\mone{\ifmmode M_1\else $M_1$\fi}
\def\tb{\ifmmode \tan\beta \else $\tan\beta$\fi}
\def\xw{\ifmmode x\subw\else $x\subw$\fi}
\def\ch{\ifmmode H^\pm \else $H^\pm$\fi}
\def\lum{\ifmmode {\cal L}\else ${\cal L}$\fi}
\def\inpb{\,{\ifmmode {\mathrm {pb}}^{-1}\else ${\mathrm {pb}}^{-1}$\fi}}
\def\infb{\,{\ifmmode {\mathrm {fb}}^{-1}\else ${\mathrm {fb}}^{-1}$\fi}}
\def\epem{\ifmmode e^+e^-\else $e^+e^-$\fi}
\def\ppb{\ifmmode \bar pp\else $\bar pp$\fi}
\def\bsg{\ifmmode B\to X_s\gamma\else $B\to X_s\gamma$\fi}
\def\bsll{\ifmmode B\to X_s\ell^+\ell^-\else $B\to X_s\ell^+\ell^-$\fi}
\def\bstt{\ifmmode B\to X_s\tau^+\tau^-\else $B\to X_s\tau^+\tau^-$\fi}
\def\lamt{\ifmmode \tilde\lambda\else $\tilde\lambda$\fi}
\def\shat{\ifmmode \hat s\else $\hat s$\fi}
\def\that{\ifmmode \hat t\else $\hat t$\fi}
\def\uhat{\ifmmode \hat u\else $\hat u$\fi}

\newskip\zatskip \zatskip=0pt plus0pt minus0pt
\def\matth{\mathsurround=0pt}
\def\lsim{\mathrel{\mathpalette\atversim<}}

\def\atversim#1#2{\lower0.7ex\vbox{\baselineskip\zatskip\lineskip\zatskip
  \lineskiplimit 0pt\ialign{$\matth#1\hfil##\hfil$\crcr#2\crcr\sim\crcr}}}

\def\grtsim{\,\,\rlap{\raise 3pt\hbox{$>$}}{\lower 3pt\hbox{$\sim$}}\,\,}
\def\lsim{\,\,\rlap{\raise 3pt\hbox{$<$}}{\lower 3pt\hbox{$\sim$}}\,\,}

\renewcommand{\thefootnote}{\fnsymbol{footnote}}

\begin{document} \begin{titlepage}
\rightline{\vbox{\halign{&#\hfil\cr
&SLAC-PUB-12191\cr
}}}
\begin{center}
\thispagestyle{empty} \flushbottom { {
\Large\bf Black Hole Production at the LHC by Standard Model Bulk Fields in the 
Randall-Sundrum Model 
\footnote{Work supported in part
by the Department of Energy, Contract DE-AC02-76SF00515}
\footnote{e-mail:
$^a$rizzo@slac.stanford.edu}}}
\medskip
\end{center}

\centerline{Thomas G. Rizzo$^{a}$}
\vspace{8pt} 
\centerline{\it Stanford Linear
Accelerator Center, 2575 Sand Hill Rd., Menlo Park, CA, 94025}

\vspace*{0.3cm}

\begin{abstract}
We consider the production of black holes at the LHC in the Randall-Sundrum(RS) model through the collisions of Standard Model(SM)  
fields in the bulk. In comparison to the previously studied case where the SM fields are all confined to the TeV brane, we find 
substantial suppressions to the corresponding collider cross sections for all initial states, \ie, $gg$, $qq$ and $gq$, where $q$ 
represents a light quark or anti-quark which lie close to the Planck brane. For $b$ quarks, which are closer to the TeV brane, 
this suppression effect is somewhat weaker though $b$ quark contributions to the cross section are already quite small  
due to their relatively small parton densities. Semi-quantitatively, we find that the overall black hole cross section is reduced 
by roughly two orders of magnitude in comparison to the traditional TeV brane localized RS model with the exact value being 
sensitive to the detailed localizations of the light SM fermions in the bulk.  
\end{abstract}



\renewcommand{\thefootnote}{\arabic{footnote}} \end{titlepage} 

%
%
%

The existence of Terascale black holes(BH) which might be produced at the LHC{\cite {BH} is a unique prediction of extra-dimensional 
models{\cite {rev}} that attempt to address the gauge hierarchy problem. In the most well-studied example, the model of Arkani-Hamed, 
Dimopoulos and Dvali(ADD){\cite {ADD}} where SM matter is confined to a brane while gravity lives in the bulk, one finds that the 
production rates for BH with masses not too far in excess of the fundamental scale can be quite large{\cite {BH}} and are 
qualitatively independent of the number of extra dimensions. This is due to the 
fact that the BH cross section at the partonic level is essentially geometric above the fundamental scale, 
$\sigma =f \pi R_s^2$, where $R_s$ is the Schwarzschild/horizon radius for a BH with mass $M_{BH}\simeq \sqrt{\hat s}$, and $f$ is a 
factor of order unity.{\footnote {To be specific in what follows we will set $f=1$ in numerical calculations.}}   
If such a scenario is naturally realized it is likely that BH production at the LHC would be observed unless we are in a very unlucky 
parameter space regime. 

In the original Randall-Sundrum(RS) model{\cite {RS}}, the SM fields were localized on the TeV brane which led to BH cross section 
estimates that were found{\cite {rsbh}} to be comparable to those obtained in the ADD case{\cite {BH}} but now assuming the existence of 
only one warped, extra dimension. More recently, for a number 
of theoretical and model-building reasons, scenarios of the RS type have been constructed wherein the SM gauge{\cite {gauge}} and 
fermion{\cite {fermion}}  fields are promoted into the bulk.{\footnote {It is also possible, with some care, 
to place the SM Higgs field in the 
bulk{\cite {bulkh}}}}. In such a situation it is no longer obvious how large the BH cross section, $\sigma_{BH}$, will 
be as one must now account, \eg, for the 5-d wavefunctions of the colliding SM zero modes. This possibility has apparently been overlooked 
in the literature and so the purpose of the present note is to examine 
this issue. We will show that placing the SM fields in the bulk generally leads to a suppression of the BH production 
rate by approximately two orders of magnitude or more in comparison to those obtained in the traditional TeV brane localized RS scenario. 

In order to make an estimate of $\sigma_{BH}$ we must make a number of assumptions; the first of these is that we can describe the region 
in the vicinity near the BH by the Schwarzschild-AdS solution{\cite {ads}} in 5-d. For this hypothesis to be valid we must assume, as is 
traditionally done, 
that the SM fields in the bulk as well as the presence of the BH itself do not significantly modify the overall background RS metric   
\begin{equation}
ds^2=e^{-2k|y|}\eta_{\mu\nu}dx^\mu dx^\nu-dy^2\,,
\end{equation}
far away from the BH, \ie, the backreaction on the global metric can be neglected. Here, $k$ describes the bulk curvature, \ie, the 5-d Ricci 
scalar in the bulk is just $R=-20k^2$. This is essentially the same type of assumption that one makes when discussing the creation 
of BH on the TeV brane in the ADD and classic RS scenarios, \ie, the BH is only a small local distortion of the global background solution 
to Einstein's equations. 

To set our conventions we note that the global bulk gravitational action is just that given by the RS model{\cite {RS}}
\begin{equation}
S_{grav}=\int d^5x ~\sqrt {-g} \Big[{M^3\over {2}} R-\Lambda_b\Big]\,,
\end{equation}
with $M$ being the fundamental scale $\sim k \sim \mpl$, the latter being the 4-d reduced Planck scale 
which satisfies $\mpl^2=M^3/k$, and $\Lambda_b=-6M^3k^2$ is 
the bulk cosmological constant inducing the AdS$_5$ curvature{\cite {RS}}. 

In performing the BH cross section calculation, we must assume that the BH is sufficiently small 
so that it does not feel the finite size associated with compactification scale $\pi r_c${\footnote {Recall that in the RS model 
there are two branes which form the boundaries of the $AdS_5$ space, one at $y=0$(Planck) and one at $y=\pi r_c$(TeV).}}. 
This is also essentially the same assumption that one makes in the case of 
matter localized on the TeV brane. A short calculation, given below, shows that for RS bulk BH with masses that are 
accessible at the LHC, this condition will always be satisfied.
Employing the Schwarzschild-AdS solution{\cite {ads}} in 5-d, one can immediately write down the relationship between the BH mass and the 
horizon radius $R_s$; one obtains{\cite {rsbh}} directly 
\begin{equation}
{\hat M_{BH}\over {3\pi^2 \hat {M^3}}}=R_s^2\big[1+(kR_s)^2\big]\,. 
\end{equation}
Taking $k/\mpl <0.1${\cite {us}}, one can easily show, as we will demonstrate below, that for BH of interest at the LHC, 
$(kR_s)^2<<1$ so that dropping this term is a very reasonable approximation. In the usual analysis of the BH formation process{\cite {yr}} 
by colliding SM particles on the TEV brane, the possible extension of the two colliding sources into the extra dimensions is already 
accounted for. Thus in the case under consideration we merely 
need to know how these colliding particles are distributed in the extra dimension, \ie, we need to know their 5-d wavefunctions. We also recall 
that since we are colliding zero mode SM fields to form the BH, there will be no momentum in the $y$ direction. 

This discussion then implies that the `5-d' parton level BH cross section is given by
\begin{equation}
\hat \sigma_{BH}=\pi R_s^2={\hat M_{BH}\over {3\pi \hat M^3}}\,. 
\end{equation}
Before going further we must ask how to interpret the values of the quantities $\hat M_{BH}$ and $\hat M$ appearing in this 
expression. Consider a BH 
sitting at some point $y$ in the bulk. As we know, the `magic' of the RS model tells us that the overall mass scale varies 
exponentially as we move between the branes, \eg, a mass $M$ on the Planck brane appears as 
$M_*=M\epsilon$, with $\epsilon=e^{-\pi kr_c}$, on the TeV brane. It is this very `running' of the mass with $y$ 
which allows us to address the hierarchy problem. One can then think of $\hat M$ and $\hat M_{BH}$ as running quantities which are 
$y$ dependent. Clearly, \eg,  the `effective' fundamental gravity scale will depend on the value of $y$ where it is being probed. 
Thus we can write the relevant mass ratio above as 
$\hat M_{BH}(y)/\hat M(y)^3=(M_{BH}|_{TeV}/M_*^3)~\epsilon^2 e^{2ky}$ where we have chosen to normalize these quantities to their TeV 
brane values to make contact with our usual 4-d picture. Hence, at any given value of $y$, we have effectively   
\begin{equation}
\hat \sigma_{BH}(y)=\epsilon^2 ~{M_{BH}|_{TeV}\over {3\pi M_*^3}}~e^{2ky}=\sigma_0 e^{2k(y-\pi r_c)}\,, 
\end{equation}
where $\sigma_0$ is the value of the BH cross section that we would obtain for matter fields localized 
on the TeV brane as in the original RS model. The result of this expression is not {\it a priori} unexpected as we know that the strength of 
gravity varies significantly between the two branes. From now on we will simply write $M_{BH}|_{TeV}=M_{BH}$. It is important to 
note that the cross section as a function of $y$ is very strongly peaked at the TeV brane. 

It is now easy to see that the `correction' term, $(kR_S)^2$, in the $M_{BH}-R_s$ relationship above arising from the fact that the space 
is asymptotically $AdS_5$, is quite small. First, note that an observer at any value of $y$ sees the same value for this dimensionless 
quantity so can we choose for convenience to evaluate it at the TeV brane. Using the well-known relationship $\mpl^2=M^3/k${\cite {RS}} 
above, we find that 
\begin{equation}
(kR_s)^2 \simeq {1\over {3\pi^2}} {k^2\over {\mpl^2}} {M_{BH}\over {k\epsilon}}\,. 
\end{equation}
However, $0.01 \leq k/\mpl \leq 0.1${\cite {us}} and $k\epsilon \sim 1$ TeV so that for any reasonable BH masses at the LHC  we obtain 
$(kR_s)^2 \lsim 10^{-4}-10^{-3}$, which can be safely neglected in comparison to unity. These light BH do not feel the global curvature 
of the $AdS_5$ space to first approximation. Note that in this limit where $(kR_S)^2<<1$, we essentially 
recover the result of the toroidally compactified ADD model assuming the existence of only one extra dimension.  
Similarly, to demonstrate that BH of interest are far smaller that the compactification scale, we can  
consider the ratio $R_s/(\pi r_c)$. Using the fact that $kr_c \simeq 11.3${\cite {us}} to explain the hierarchy and the value of 
$kR_s$ that we just obtained above, we then find that 
$R_s/(\pi r_c)=kR_s/(11.3\pi) \lsim 10^{-3}$. Thus the BH are quite small and we can also safely ignore effects due to the finite size 
of the compactified dimension in our calculations.

Now let us calculate the 4-d partonic level cross section due to the collision of two SM 5-d fields $F_{1,2}$ which have 
(suitably normalized) 5-d wavefunctions $f_i(y)$; one obtains
\begin{equation}
(\sigma_{BH})_{12}=\int_{-\pi r_c}^{\pi r_c}~dy~\hat \sigma_{BH}(y)f_1(y)f_2(y)\,. 
\end{equation}
Of course it is understood that there is also a $\theta$ function here as well which tells us, as is the usual prejudice, that BH are 
not produced unless the collision energies (as measured on the TeV brane) exceed $M_*$. 
Since SM fields are distributed in the bulk, one can think of the integrand in this expression as the probability of forming a BH at 
any particular point in the extra dimension which we then need to integrate over. Clearly the probability is largest near the TeV 
brane where the effective gravitational mass scale is the smallest. Naively, for any given $F_{1,2}$, we expect that the SM fields which are 
peaked closest to the TeV brane will give the largest contributions to the production cross section before being weighted by the 
appropriate parton densities. 

As a first and most simple 
application of our analysis, we consider the process $gg\to $BH, where $g$ is the bulk zero-mode gluon field. Since the gluons are 
massless gauge fields, in the absence of brane terms{\cite {brane}}, $f_{1,2}=(2\pi r_c)^{-1/2}$ so that the above 
integral is trivial; we thus obtain the partonic cross section 
\begin{equation}
\sigma_{BH,gg}=S_{gg}\sigma_0={{\sigma_0}\over{2\pi kr_c}}\simeq {1\over {71}}\sigma_0\,, 
\end{equation}
where we have used the fact that $kr_c \simeq 11.3${\cite {us}} to deal with the hierarchy problem and have dropped terms of relative 
order $\epsilon^2$. This result is shown in Fig~\ref{fig1}. Note that by placing the gluons in the bulk the effective cross section 
is reduced in comparison to gluons localized on the TeV brane by the suppression factor, $S_{gg}$, which is almost two orders of 
magnitude. This result is {\it independent} of where the SM 
fermions are localized in the bulk and will thus be valid in all models with bulk gauge fields. Since the $gg$ luminosity is very high at 
the LHC the reduction in the cross section for this subprocess is extremely important and is essentially model independent. 
\begin{figure}[htbp]
\centerline{
\includegraphics[width=7.5cm,angle=90]{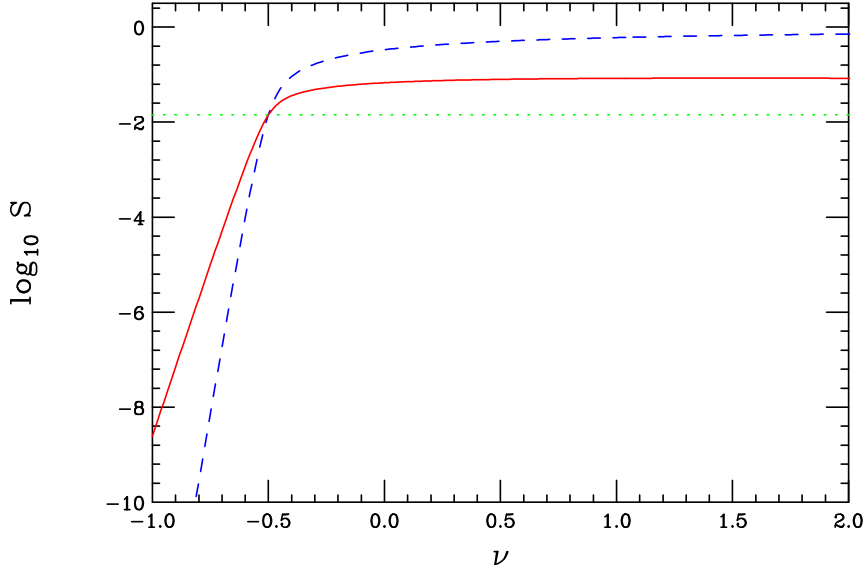}}
\vspace*{0.1cm}
\caption{Suppression factors $S_{ff,gf}$ as functions of $\nu$ for fermion pair(blue dashed) and glue-fermion(solid red) production 
of BH in the RS bulk, respectively. The constant value of $S_{gg}$(green dots) is shown for comparison purposes.}
\label{fig1}
\end{figure}

Now let us turn to the case of the process $ff'\to$ BH, where $f,f'$ are two zero mode fermions; in practical applications these 
fermions will be left- or right-handed quarks or anti-quarks. Such fields have 5-d wavefunctions $\sim e^{(\nu+1/2)k|y|}$, 
where we follow the notation of our earlier work{\cite {fermion}} with the value of the parameter $\nu$ being determined by the 
5-d fermion bulk mass, \ie, a term of the form $\sim -(\rm{sign} ~y) k\nu$ in the integrand of the action.{\footnote {The parameter 
$c=-\nu$ is often used instead in the literature{\cite {fermion}}.}}. 
For $\nu >(<)-1/2$, the fermion is seen to be localized near the TeV (Planck) brane. 
Performing the integrations in a straightforward manner(after properly normalizing the fermion wavefunctions), we obtain 
\begin{equation}
\sigma_{BH,ff'}=S_{ff'}\sigma_0=\sigma_0{{(1+2\nu)^{1/2}(1+2\nu')^{1/2}}\over {\nu+\nu'+3}}
~{{e^{\pi kr_c(1+\nu+\nu')}-\epsilon^2}\over {N(\nu)N(\nu')}}\,, 
\end{equation}
where the explicit normalization factors are given by 
\begin{equation}
N(\nu)=\Big[e^{\pi kr_c(1+2\nu)}-1\Big]^{1/2}\,. 
\end{equation}

To see how large the resulting cross section suppression can be, let us first consider the simpler case where  $\nu=\nu'$; here we find 
that 
\begin{equation}
S_{ff}={{1+2\nu}\over {3+2\nu}}~{{e^{\pi kr_c(1+2\nu)}-\epsilon^2}\over {e^{\pi kr_c(1+2\nu)}-1}}\,, 
\end{equation}
which is shown in Fig.~\ref{fig1}. 
For fermions localized very near the TeV brane, $\nu$ is large and positive and $S_{ff}\sim (2\nu+1)/(2\nu+3)$ 
is only somewhat smaller than unity. (Note that as $\nu \to \infty$, \ie, matter localized on the TeV brane, $S_{ff}\to 1$ 
as it should.) As 
$\nu$ decreases the effect of the suppression increases. For example, at $\nu=-1/2$ we find the same suppression as in the glue-glue case, 
$S_{ff}=1/(2\pi kr_c)$, since in this example the fermion wavefunctions would then be flat in $y$. 
After this point, as $\nu$ becomes more negative, the effect of $S_{ff}$ becomes exponentially strong  and the cross section 
rapidly becomes effectively zero for all practical purposes. Thus for $qq \to$BH type processes we must have the quarks as close to the 
TeV brane as possible for them to make a reasonable contribution to the cross section and to avoid any huge suppression factors. 

\begin{figure}[htbp]
\centerline{
\includegraphics[width=7.5cm,angle=90]{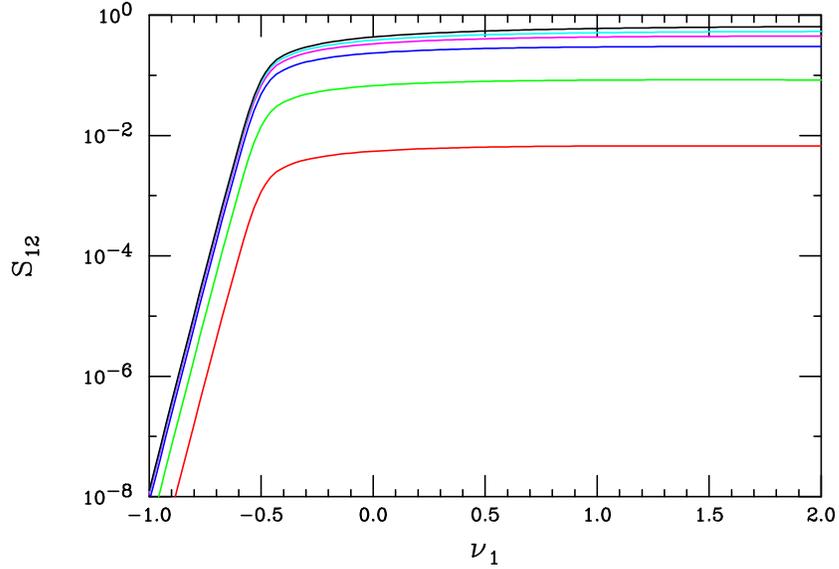}}
\vspace*{0.1cm}
\caption{Fermion pair suppression factors $S_{ff'}$ as functions of $\nu_1$ for (from bottom to top) $\nu_2=-0.6, -0.5, -0.3, 0.0, 0.3$ and 
1.0, respectively.}
\label{fig0}
\end{figure}
In the more general case of two different fermions types, we obtain the results shown in Fig.~\ref{fig0}. Here 
there are several factors acting simultaneously 
including the relative separation of the two fermions in the bulk as well as the overall peaking structure of the 5-d cross section. 
We will need these general results when calculating the total BH cross section at the LHC.

A final case to consider is the process $gf\to$ BH. Following the by now standard calculation we arrive at the suppression factor for 
this reaction given by 
\begin{equation}
S_{gf}={1\over {\sqrt {\pi kr_c}}}~{{\sqrt {1+2\nu}}\over {\nu+5/2}}~{{e^{\pi kr_c(\nu+1/2)}-\epsilon^2}\over 
{[e^{\pi kr_c(2\nu+1)}-1]^{1/2}}}\,, 
\end{equation}
which is also shown in Fig.~\ref{fig1}. 
Here we see that for large positive $\nu$, $S_{gf} \simeq (2\pi kr_c)^{-1/2} \simeq 1/8.4$ so that there is 
already a sizable suppression. As $\nu$ decreases further this suppression grows slowly and, as expected, reaches 
$S_{gf}=1/(2\pi kr_c)$ when $\nu=-1/2$ 
as in the glue-glue case. For more negative values of $\nu$, we again see an exponential suppression with, roughly, 
$S_{gf}\simeq S_{ff}^{1/2}$ in this region. Thus for reactions of this class to be important the relevant fermion must again 
lie as close as possible to the TeV brane just as we observed in the previous $ff'\to$ BH case. 

\begin{figure}[htbp]
\centerline{
\includegraphics[width=8.5cm,angle=90]{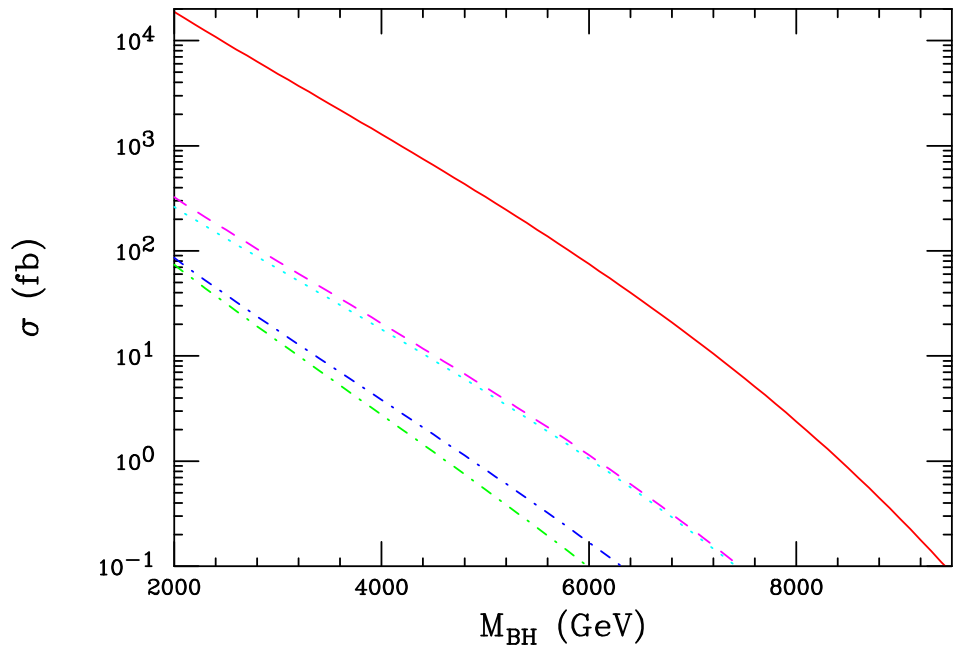}}
\vspace*{0.1cm}
\caption{BH production cross section at the LHC assuming $M_*=2$ TeV for purposes of demonstration. The top(solid red) curve is for the 
familiar case where SM fields are 
localized on the TeV brane. The other curves, as discussed in the text, correspond to gluons in the bulk with the SM fermions localized in 
the bulk in various ways.}
\label{fig2}
\end{figure}

What are the implications of all these suppression factors for actual BH production at the LHC? Here we will consider a few sample cases to 
get a feel for the variety of possibilities allowed by the model parameter space. To this end we need to examine the relevant ranges  
for the $\nu$ parameters for light (\ie, first and second generation) quarks as well as for 
$b_{L,R}$. As we will see, although the $b_{L,R}$ are closer to the TeV brane than the lighter fields in order to generate their larger 
mass, there is still a sizable suppression in this case and the small value for the 
relevant parton densities makes their overall contribution to the BH cross section quite small. 
Clearly top quarks can make no numerically relevant contribution to the BH cross section even if they were localized to the SM brane.  

Several sets of authors have attempted realistic RS models with bulk localized SM fermions{\cite {bulkf}}. As a first example, 
we consider localizing the light first and second generation quarks with $\nu=-0.5$ and $b_L(b_R)$ with $\nu=-0.3(-0.4)$, respectively. 
(Occasionally, more negative values of $\nu$ are chosen for the light quarks{\cite {bulkf}}, a situation we will return to below.) 
With this parameter choice, \eg, the couplings of these zero mode fermions to the KK towers of the SM gauge fields are quite small so as 
to not upset precision electroweak fits{\cite {bulkf}}.
Using the equations and assumptions above we can immediately calculate the expected BH cross section at the LHC. The results of this 
analysis are shown as the dashed magenta curve 
in Fig.~\ref{fig2}, assuming for demonstration purposes that $M_*=2$ TeV.{\footnote {It is important 
to note that $M_*$ is {\it not} the same as the parameter $\Lambda_\pi$ usually employed in RS phenomenological analyses{\cite {us}}. In 
fact, one has $\Lambda_\pi^2=M_*^3/k\epsilon$.}}  These are shown together with the 
results in the traditional RS case (the solid red curve at the top of the figure) where all of the SM particles are localized on 
the TeV brane for comparison purposes.{\footnote {Note that for other values of $M_*$ we can simply use the $M_*^{-3}$ scaling behavior 
demonstrated above to obtain the desired cross section, remembering also the assumed BH threshold at $\sqrt {\hat s}=M_*$.}} In this case, 
in comparison to the TeV localized SM particle scenario, the cross section is seen to be reduced by a factor of 
$\simeq 58(66)$ for a BH of mass 2(6) TeV, respectively. This is {\it very} roughly the same effect as increasing 
$M_*$ to $\simeq 8$ TeV with all the SM fields on the TeV brane, 
except that the scaling is only approximate and that the threshold of the cross 
section still remains at 2 TeV. Since the overall cross section is {\it not} 
simply rescaled and given the fact that the threshold remains low, it should be clear from the BH production data at the LHC that 
there is indeed a suppression effect acting in comparison to TeV localization expectations. 

How significant are the $b_{L,R}$ 
contributions to this cross section? To address this issue, we leave the light fermions where they are and simply set to zero the 
contributions arising from $b$ quarks; from this we obtain the dotted cyan curve in Fig.~\ref{fig2}. 
In this case, for both $M_{BH}=2,6$ TeV we get a 
suppression of the cross section by a factor of $\simeq 71$, which is similar to assuming $M_*\simeq 8.3$ TeV for TeV brane localized fields, 
except for the threshold 
effect. In this case we are essentially seeing a uniform suppression of the $gg, gq$ and $qq$ cross sections by an identical factor. 

How sensitive are our results to the assumed location of the light fermions?  Moving the light fermions away from $\nu=-0.5$ will result in 
different suppressions in each of the $gg, gq$ and $qq$ channels. To address this, we first restore the $b_{L,R}$ quarks to 
their former locations and then move the light quarks to $\nu=-0.6(-0.7)$; the results are given by the dash-dotted (upper)blue and 
(lower)green curves in Fig.~\ref{fig2}. For such values of $\nu$, we are apparently almost completely turning 
off the $gq \to $BH and $qq\to $BH processes. We now observe a very large suppression of the BH cross section by roughly, in the case 
of $\nu=-0.6$, a factor of 220(445) for BH of mass 2(6) TeV, respectively, which does not show any effective $M_*$-like scaling behavior. 
For $\nu=-0.7$, the observed suppression is somewhat larger by about a factor of 1.2-2 depending upon the BH mass. 

Although an analysis is beyond the scope of the present paper, we would like to comment about how 5-d localization of the SM fields 
may modify the usual BH emission probabilities and corresponding BH lifetimes. Clearly, the distributions of the SM fields in the bulk 
will lead to suppressions in BH decays that will be felt through modifications of the usual greybody factors{\cite {BH}}. Recall that 
these are usually obtained by matching the wavefunction for a particle of a given spin near the BH horizon to that obtained in the 
asymptotic region far away from the BH that are usually calculated in D-dimensional Schwarzschild co-ordinates. This standard approach 
will not work 
in our case since the known 5-d parts of the asymptotic wavefunctions for various fields are not those obtained from 
Schwarzschild co-ordinates.  Close to the BH and at intermediate distance it would perhaps be more appropriate to employ the 
co-ordinates of the Schwarzschild-AdS solution{\cite {ads}}. Asymptotically, we know the solutions are of the pure $AdS_5$ form and these 
two sets would then need to be matched at intermediate distances. Such calculations have as of yet not been performed in the literature 
but one might anticipate some suppression of decay rates leading to somewhat longer BH lifetimes. How large an effect these may be would 
require a detailed calculation.

Before closing we note that the BH cross section can also be suppressed in non-warped models where fermions are localized at different points 
in the bulk as in split-fermion models{\cite {split}}. However, in these cases one fines that the overall suppression factors in such cases 
are only of order unity{\cite {stoj}} unlike the two order of magnitude effects discussed here.

Summarizing, we have investigated the relative suppression of the BH production cross section at the LHC in the RS model when the SM fields 
are placed in the bulk. Depending upon the BH mass and the locations of the various fermions we have found suppression factors lying in 
the range $\simeq 60-500$. Although this significantly reduces the volume of the parameter space region region over which BH may be 
observed at the LHC, a reasonable fraction of this region remains accessible. If BH are observed at the LHC it is likely that the 
suppression factors discussed here can be measured provided the BH are realized within the RS context. For the future we must address the 
issue of the appropriate greybody factors for BH decay into the SM fields in the $AdS_5$ bulk.

\noindent{\Large\bf Acknowledgments}

The author would like to thank J. Hewett for discussions related to this paper.

%
\def\MPL #1 #2 #3 {Mod. Phys. Lett. {\bf#1},\ #2 (#3)}
\def\NPB #1 #2 #3 {Nucl. Phys. {\bf#1},\ #2 (#3)}
\def\PLB #1 #2 #3 {Phys. Lett. {\bf#1},\ #2 (#3)}
\def\PR #1 #2 #3 {Phys. Rep. {\bf#1},\ #2 (#3)}
\def\PRD #1 #2 #3 {Phys. Rev. {\bf#1},\ #2 (#3)}
\def\PRL #1 #2 #3 {Phys. Rev. Lett. {\bf#1},\ #2 (#3)}
\def\RMP #1 #2 #3 {Rev. Mod. Phys. {\bf#1},\ #2 (#3)}
\def\NIM #1 #2 #3 {Nuc. Inst. Meth. {\bf#1},\ #2 (#3)}
\def\ZPC #1 #2 #3 {Z. Phys. {\bf#1},\ #2 (#3)}
\def\EJPC #1 #2 #3 {E. Phys. J. {\bf#1},\ #2 (#3)}
\def\IJMP #1 #2 #3 {Int. J. Mod. Phys. {\bf#1},\ #2 (#3)}
\def\JHEP #1 #2 #3 {J. High En. Phys. {\bf#1},\ #2 (#3)}

\end{document}